\documentclass[prl, twocolumn, amsmath, amssymb,superscriptaddress]{revtex4}
\usepackage{graphics,graphicx,color,amsmath,amssymb,bm,hyperref,subfigure}

\usepackage{bm,dsfont}

\begin{document}

\definecolor{Blue}{rgb}{0,0,1}
\definecolor{Red}{rgb}{1,0,0}
\definecolor{Green}{rgb}{0,1,0}
\definecolor{Purp}{rgb}{.2,0,.2}
\definecolor{white}{rgb}{1,1,1}
\newcommand{\rev}[1]{{\color{Red} #1}}

\title{Thermodynamics of quantum coherence}

\author{C\'{e}sar A. Rodr\'{i}guez-Rosario$^{1}$, Thomas Frauenheim$^{1}$  Al\'{a}n Aspuru-Guzik$^2$\\ \small
$^1$ Bremen Center for Computational Materials Science, University of Bremen, Am Fallturm 1, D-28359, Bremen, Germany\\ \small
$^2$ Department of Chemistry and Chemical Biology, Harvard
  University, Cambridge, Massachusetts 02138, USA}

%



\begin{abstract}
Quantum decoherence is seen as an undesired source of irreversibility that destroys quantum resources~\cite{Zurek:2003uu}. Quantum coherences seem to be a property that vanishes at thermodynamic equilibrium. Away from equilibrium, quantum coherences challenge the classical notions of a thermodynamic bath in a Carnot engines~\cite{Scully:2003vt,Scully:2010cn}, affect the efficiency of quantum transport~\cite{Mohseni:2008wt,Plenio:2008ff,Rebentrost:2009tj}, lead to violations of Fourier's law~\cite{Manzano:2011wz}, and can be used to dynamically control the temperature of a state~\cite{Erez:2008ep}. However, the role of quantum coherence in thermodynamics~\cite{Prigogine98} is not fully understood. Here we show that the relative entropy of a state with quantum coherence with respect to its decohered state captures its deviation from thermodynamic equilibrium. As a result, changes in quantum coherence can lead to a heat flow with no associated temperature, and affect the entropy production rate~\cite{Spohn:1978uv}. From this, we derive a quantum version of the Onsager reciprocal relations \cite{Onsager:1931uu} that shows that there is a reciprocal relation between thermodynamic forces from coherence and quantum transport. Quantum decoherence can be useful and offers new possibilities of thermodynamic control for quantum transport \cite{Solomon:2008ej,Vazquez:2012gg}.
\end{abstract}

\maketitle

\subsection{Introduction}
 The evolution of a system state $\rho$ can be described by a master equation $\dot{\rho}=-i\left[H,\rho \right]+\mathbb{L}\left(\rho\right)$ where $H$ is the Hamiltonian of the system, and $\mathbb{L}$ describes the coupling to a Markovian bath~\cite{Gorini:1976vs}. The solution of this equation is the dynamical map $\rho(t)=\mathbb{B}^{(0,t)}\left(\,\rho(0)\, \right)$~\cite{Sudarshan:1961ua}. Determining $\mathbb{L}$ is experimentally demanding, requiring quantum process tomography~\cite{Nielsen00a}. To overcome this difficulty, we will focus instead on quantum thermodynamic properties that depend on equilibrium and deviations from it. We provide a description of the role of decoherence in terms of the change in energy, entropy and entropy production. Finally, we introduce a quantum version of the Onsager reciprocal relations between decoherence and transport.

To understand the thermodynamics of quantum coherence, we must go beyond characterizing thermodynamic equilibrium simply by a temperature parameter. As our starting point, we consider the stationary states of a quantum process. These are reached when a system is coupled to a bath for long enough such that $\rho\rightarrow\mathbb{B}\left(\rho \right)\equiv\lim_{t \to \infty}\mathbb{B}^{(0,t)}\left(\rho \right)$. The state $\mathbb{B}\left(\rho \right)$ is stationary because  $\mathbb{L}\left(\,\mathbb{B}\left(\rho \right)\,\right)=0$~\cite{Lindblad83}. All stationary states $\{\eta\}$ with the property that $\mathbb{L}\left(\eta\right)=0$ (or equivalently $\mathbb{B}\left(\eta \right)=\eta$) form the stationary set. We propose to use the stationary set of the quantum process as the way to characterize a quantum thermodynamic bath. This captures classical thermodynamics for the case of a relaxation process (subscript $r$) where any state $\rho$ evolves as $\dot{\rho}=\mathbb{L}_r\left(\rho\right)$ becoming the Gibbs stationary state $\rho\rightarrow \mathbb{B}_{r}\left(\rho \right)=\frac{e^{-\beta H}}{\mbox{Tr}\left(e^{-\beta H}\right)}$. However, focusing on stationary sets also allows us to consider decoherence as a new thermodynamic process. Decoherence is a quantum process that, by means of a master equation $\mathbb{L}_d$, destroys the off-diagonal elements (coherences) of a density matrix. The stationary set of decoherence can be described with a set of classical probability vectors on the preferred basis, $\rho\rightarrow\mathbb{B}_{d}\left(\rho\right)=\sum_j \vert j \rangle\langle j \vert  \rho \vert j \rangle\langle j \vert=\sum_j p_j\vert j \rangle\langle j \vert$. A decoherence bath can be characterized simply by the preferred basis $\{\vert j \rangle\}$ of the stationary set. 

\subsection{Zeroth law of thermodynamics}

\begin{figure}[t!]\begin{center}
{\includegraphics[keepaspectratio=true,width=\linewidth]{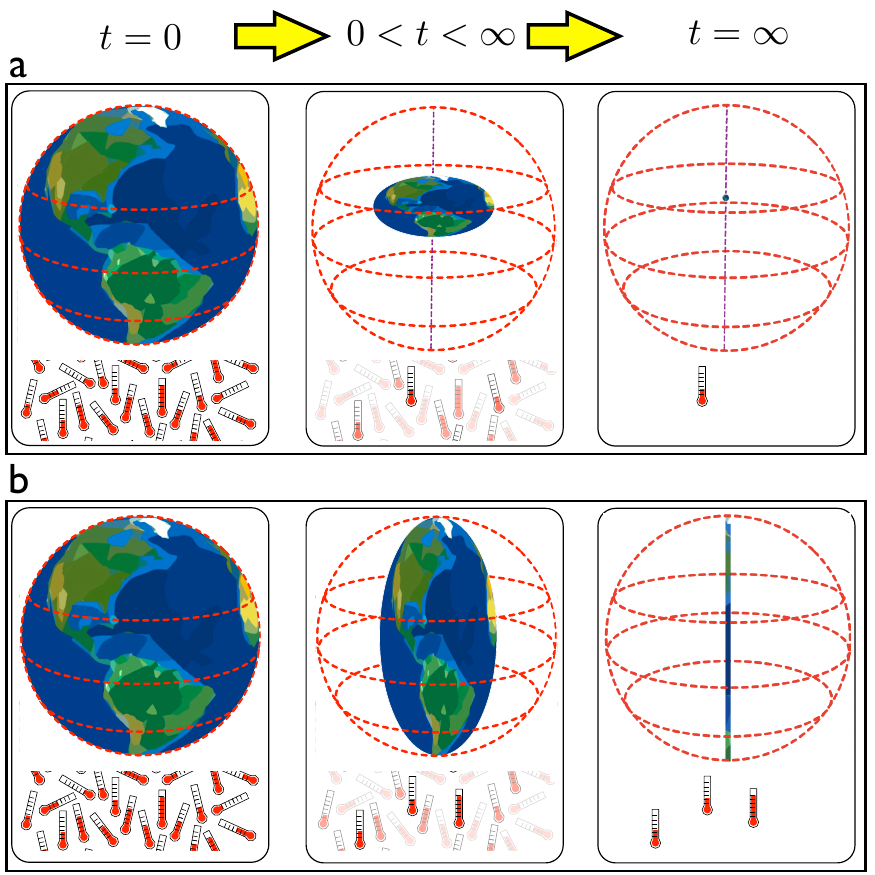} }
\end{center}
\caption{\emph{Approach to equilibrium.}  The solid ball painted as Earth represents the Bloch sphere of possible density matrices for a two-level system. Cartoon thermometers represent equilibrium. {\bf a}, A relaxation bath squeezes all possible states into a unique point that corresponds to the  Gibbs state. A `thermometer' state would contain information about the system Hamiltonian and the unique temperature of this bath. {\bf b}, A decoherence bath squeezes the states into a line along the preferred basis that corresponds to the set of states without quantum coherences. Any `thermometer' state in the preferred basis is in equilibrium, but the temperature is not uniquely defined.}\label{fig:sphere}
\end{figure}

The zeroth law of thermodynamics is a statement about how systems can act like `thermometers' such that they are stationary upon coupling to a bath. To quantify the \emph{surprise} of a system with respect to a bath, we will use the concept of relative entropy\cite{Vedral:2002vd}. The relative entropy of a system $\rho$ with respect to its corresponding stationary state for process $\mathbb{B}$ is
\begin{equation}\label{eq:relativenetropy}
\mathbf{R}\left[\rho \,\Vert \,\mathbb{B}\, \right]=\mbox{Tr}\left[\rho \log \rho \right]-\mbox{Tr}\left[\rho \log \mathbb{B}\left(\rho\right) \right],
\end{equation}
This quantity captures that the state $\rho$ is far from the stationary set of the process $\mathbb{B}$. We now express the zeroth law of quantum thermodynamics as: a state $\rho$ is in quantum thermodynamic equilibrium with a bath $\mathbb{B}$ when
 $\mathbf{R}\left[\rho \,\Vert \,\mathbb{B}\, \right]=0$. Although the states on quantum thermodynamic equilibrium might not be unique, the system acts like a thermometer in the sense that there is no surprise from being coupled to the bath. Since  $\mathbf{R}\left[\rho \,\Vert \,\mathbb{B}\right]$ cannot increase in time (see Appendix A), a state in contact with a bath tends to evolve towards quantum thermodynamic equilibrium.   
  When considering relaxation to a Gibbs state, this fully captures classical thermodynamics. For relaxation dynamics, the Gibbs state is the only stationary state and can be described solely in terms of the temperature $\beta$ and the system Hamiltonian. Any change in energy or temperature of the system will take it away from thermodynamic equilibrium~ (see Fig.~\ref{fig:sphere}a) with respect to the Gibbs state \cite{Prigogine98}. 
  
  The definition of equilibrium given by Eq.~(\ref{eq:relativenetropy}) goes beyond the Gibbs formalism. We now use the zeroth law to look at decoherence. The set of decohered states $\{\eta_d=\sum_j p_j\vert j \rangle\langle j \vert\}$ for all possible probability vectors $\{p_j\}$ are in quantum thermodynamic equilibrium with respect to decoherence because $\mathbf{R}\left[\eta_d \,\Vert \,\mathbb{B}_d\, \right]=0$. Equilibrium of a decoherence process is not given by a unique Gibbs state, but by the entirety of all states in the preferred basis. For decoherence, a quantum state acts as a `coherence thermometer' when it has lost its coherence such that it is not surprised by the (classical) stationary distribution. The information that it obtains from the decoherence bath is the preferred basis (see Fig.~\ref{fig:sphere}b). 

\subsection{First law of thermodynamics}

A decoherence bath can also create a heat flow. To show this, we start with the first law of quantum thermodynamics, $\frac{d}{dt}\mathbf{E}=\dot{\mathbf{W}}+\dot{\mathbf{Q}}$, that expresses the change of energy 
$\frac{d}{dt}\mathbf{E}=\mbox{Tr}\left[H\rho\right]$ in terms of the work rate $\dot{\mathbf{W}}=\mbox{Tr}[ \dot{H}\rho ]$ and the heat rate $\dot{\mathbf{Q}}=\mbox{Tr}\left[ H\,\mathbb{L}(\rho)\right]$. Coupling to a relaxation bath $\mathbb{L}_r$ recovers classical thermodynamics\cite{Alicki:1979uo}. To go beyond this, we instead consider the decoherence operator $\mathbb{L}_d$ and the Hamiltonian $H(\tau)=\sum_k E_k \vert E(\tau)_k \rangle\langle E(\tau)_k\vert$  that at time $\tau$ it is controlled such that it is not on the preferred basis. Then, the system coherence leads to an energy exchange with the decoherence bath. This heat rate is driven by the change of quantum coherence due to the decoherence bath, a bath for which no temperature can be defined. Such heat rates driven by decoherence can be used to define a quantum Carnot engine whose efficiency depends on coherence~\cite{Scully:2003vt,Scully:2010cn} and for temperature control~\cite{Erez:2008ep} (see Appendix C).

\subsection{Second law of thermodynamics} 

Irreversibility due to decoherence also plays an important role. To study it, we take the time derivative of Eq.~(\ref{eq:relativenetropy}) to obtain the entropy rate equation,
\begin{equation}\label{eq:entropyrate}\begin{array}{ccc}
\underbrace{-\mbox{Tr}\left[\dot{\rho} \log \rho \right]}&=  \underbrace{\mbox{Tr}\left[\dot{\rho} \log \mathbb{B}\left(\rho\right) \right]} & \underbrace{- \frac{d}{dt}\mathbf{R}\left[ \rho\Vert\mathbb{B}\right]},\\
\dot{\mathbf{S}}=&  -\boldsymbol{\Phi} & +\mathbf{P},
\end{array} 
\end{equation}
where $\mathbf{S}=-\mbox{Tr}\left[\rho \log \rho \right]$ is the von Neumann entropy of the system, $\boldsymbol{\Phi}$ is the entropy flux due to the bath, and $\mathbf{P}$ is the entropy production rate. The second law of quantum thermodynamics can be written therefore as $\mathbf{P}=- \frac{d}{dt}\mathbf{R}\left[ \rho\Vert \mathbb{B}\right]\geq0$, which states that irreversibility cannot decrease the quantum entropy production (see Appendix B). Classical non-equilibrium thermodynamics corresponds to the special case of a relaxation bath $\mathbb{B}_r$~\cite{Spohn:1978uv}. Decoherence baths not only follow the second law~\cite{haenggi:1cv,Brandao:2013vda}, but also provide an additional quantum contribution to entropy production.

%

\subsection{Onsager reciprocal relations}
Now, we examine the implications of an additional source of irreversibility due to decoherence. This corresponds to a scenario where the system is coupled to many baths $\{b\}$, each with an operator $\mathbb{L}_b$. Independently, each bath has its own, different, stationary set in $\{\eta_b \}$ in thermodynamic equilibrium, $\mathbf{R}\left[\eta_b \,\Vert \,\mathbb{B}_b\, \right]=0$. However, the interplay between all the baths keeps the system in a non-equilibrium steady state $\nu$ that is not in thermodynamic equilibrium with any of the baths~\cite{Mazur}. The entropy flux to each bath is then $\boldsymbol{\Phi}_b=-\mbox{Tr}\left[\mathbb{L}_b(\nu) \log \mathbb{B}_b\left({\nu}\right) \right]$. It follows that the entropy production rate from the interplay amongst multiple baths is
%
\begin{align}\label{eq:ness}
\mathbf{P}  &= \sum_b\mbox{Tr}\left[\mathbb{L}_b(\nu)\left(\log \mathbb{B}_b\left(\nu\right) - \log \nu\right)\right],\nonumber\\
&=\sum_b\mbox{Tr}\left[J_b\; X_b \right],
\end{align}
where  $J_b= \mathbb{L}_b(\nu)$ represents a flow and $X_b=\log \mathbb{B}_b\left(\nu\right) - \log \nu$  is the corresponding thermodynamic force. This is the quantum generalization of the entropy production rate for a non-equilibrium steady state. The rate of change represented by $J_b$ says how, away from equilibrium, there is a flow to bath $b$. The force $X_b$ represents how far the non-equilibrium steady state is from the stationary set for bath $b$. Since the interplay between coherences in the steady state and decoherence can lead to a thermodynamic force $X_d$, decoherence plays an important role in the entropy production even under the presence of other (classical) relaxation baths.

The discovery of the Onsager reciprocal relations were a turning point in thermodynamics by providing general nonequilibrium results that applied without any specific details of the model studied\cite{Onsager:1931uu}. We now derive more general quantum relations that can be applied to study the nonequilibrium role of coherence. The use of Eq.~(\ref{eq:ness}) requires knowledge of the details of the non-equilibrium dynamics of each bath $\mathbb{L}_b$. To simplify this, we approximate the current linearly in terms of the forces~\cite{Onsager:1931uu}. Since $J_b$ and $X_b$ are matrices, the linearization corresponds to a super-operator $\mathbb{M}_{b,a}$ acting on the forces: $J_b\approx \sum_a \mathbb{M}_{b,a}(X_a)$. In this regime, the quantum entropy productions can be written in terms of the forces as:
\begin{equation}\label{eq:onsager}
\mathbf{P}=  \sum_{a,b}\mbox{Tr}\big[\mathbb{M}_{b,a}(X_a) \, X_b \big]=\sum_{a,b}\mbox{Tr}\big[\mathbb{M}^\dagger_{a,b}(X_b) \, X_a \big],
\end{equation}
with the quantum reciprocal relations $\mathbb{M}_{b,a}(\,\cdot\,)=\mathbb{M}_{a,b}^\dagger(\,\cdot\,)$ (see Appendix D). The quantum reciprocal relations give us a phenomenological way to understand the interplay between different quantum baths in terms of the deviations from equilibrium with respect to each bath. The quantum reciprocal relations apply for any quantum Markovian bath and give a relationship that is independent of the specific microscopic details of the dynamics. We now use this to show how quantum coherence is a thermodynamic force in quantum transport.

\begin{figure}[t!]
\centering
\includegraphics[keepaspectratio=true,width=\linewidth]{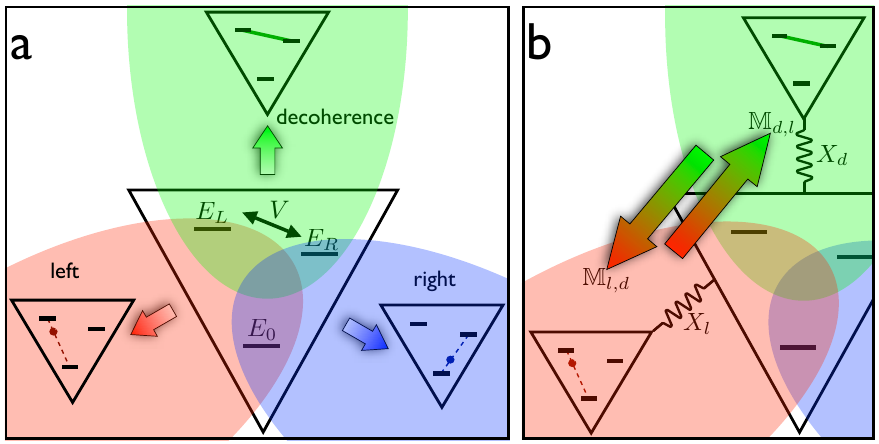}
\caption{\emph{ Interplay of transport and decoherence.} {\bf a,} A device used for quantum transport is represented by the three energy levels. The interplay between the left and the right relaxation baths leads to a flow that, with the coupling $V$, creates a coherence $\langle E_L \vert V \vert E_R\rangle$. The decoherence bath creates a flow that destroys this coherence. {\bf b,} The interplay between the baths can be approximated using quantum reciprocal relations. The left bath creates a force  (spring $X_l$) that pulls the device towards the stationary Gibbs state. The decoherence bath creates a force (spring $X_d$) that pulls the device towards a state with no coherence. This in turn create flows (arrows $\mathbb{M}_{l,d}$ and $\mathbb{M}_{d,l}$) that are reciprocally related. }  \label{fig:thermodephasing}
\end{figure}

\subsection{Thermodynamic role of coherence in transport}
Coherence can have an enhancing effect on quantum transport. It helps in energy transfer in photosynthetic complexes~\cite{Mohseni:2008wt, Plenio:2008ff, Rebentrost:2009tj} and affects quantum transport in nanoscale devices~\cite{Solomon:2008ej,Vazquez:2012gg,Segal:2001ud}. These suggest that there is a general effect, which we now explain using the quantum reciprocal relations. We construct a simple model that has all the essential features by considering a central quantum system, the device, between two relaxation baths $L$ and $R$. The system has a Hamiltonian term $V$ that can create quantum coherence between the left and the right parts of the device. Although the baths are classical, the non-equilibrium steady state sustains a quantum coherence. Quantum transport from one bath to the other through the device is mediated in part by this quantum coherence. Decoherence can be seen as an additional bath $D$ that changes the non-equilibrium steady state, and in turn, the quantum transport (see Fig.~\ref{fig:thermodephasing}). Using Eq.~(\ref{eq:onsager}) we conclude that the flow of quantum coherence into the decoherence bath has a reciprocal relation with the quantum transport between $L$ and $R$. The coherence coming from the flow through the device affects the amount of decoherence. Reciprocally, the amount of decoherence affects the quantum transport between $L$ and $R$ (see Appendix E). This effect could be experimentally verified on a molecular junction~\cite{Vazquez:2012gg}
 by controlling decoherence~\cite{Xu:2013hj}.

\subsection{Conclusions}

We have shown how quantum coherences lead to new thermodynamic flows and forces. For this, we defined quantum equilibrium in terms of relative entropy. This permitted us to write the laws of quantum thermodynamics in a way that we can apply them to study decoherence. We showed how decoherence can lead to heat flows and change the entropy production. We used these to generalize the Onsager relations to the quantum regime, which lead to a simple explanation of the role of coherence in quantum transport. This work prompts further studies on how to use coherence as a thermodynamic resource, such as in the most recent experimental studies on heat dissipation in atomic-scale junctions \cite{Lee:2013fc} and in experiments that use thermodynamic baths to created quantum information resources at steady-state \cite{Lin:2013te}.

\subsection{Acknowledgements} This work (A.A.-G. and C.A.R.R.) was supported by the Center of Excitonics, an Energy Frontier Research Center funded by the US Department of Energy, Office of Science, Office of Basic Energy Sciences under Award Number DESC0001088. A.A.-G. also thanks the Corning foundation for their generous support. C.A.R.R. thanks Peter Love, Stephanie Wehner and Kavan Modi for helpful discussions.
%

\section{Appendices}

\subsection{A. Relative entropy and equilibrium}

The relative entropy of the density matrix $\rho$ with respect to $\sigma$ is defined as
\begin{equation}
S\left[ \,\rho\, \Vert\, \sigma\,\right]=\mbox{Tr}\left[\rho \log \rho \right]-\mbox{Tr}\left[\rho \log \sigma \right].
\end{equation}
This expression characterizes the surprise of gaining the state $\sigma$ when having the state $\rho$ and is a measure of the information loss when trying to approximate $\sigma$ with $\rho$. For further details on the relative entropy, we refer the reader to the review by Vedral~\cite{Vedral:2002vd}. Relative entropy is well-defined for states that are not pure, but there are techniques~\cite{Audenaert:2011vm} that can be applied to overcome this limitation~\footnote{For simplicity, here we consider that the $\log$ is taken over the support of the matrices.}. Relative entropy never increases when the states evolve under the dynamics of a completely positive map $\mathbb{A}$, such that 
\begin{equation}\label{rentropydec}
S\left[ \,\rho\, \Vert\, \sigma\,\right]\geq S\left[\,\mathbb{A}^{(0,t)}(\rho)\, \Vert\, \mathbb{A}^{(0,t)}(\sigma)\,\right].
\end{equation}

Here we consider the special case of the relative entropy of a state $\rho$ with respect to its stationary state $\lim_{t\rightarrow\infty}\mathbb{B}^{(0,t)}(\rho)=\mathbb{B}(\rho)$:
\begin{equation}\label{eq:relativenetropy1}
\mathbf{R}\left[\rho \,\Vert \,\mathbb{B}\, \right]=S\left[\,\rho\, \Vert\, \mathbb{B}(\rho)\,\right]=\mbox{Tr}\left[\rho \left(\, \log \rho - \log \mathbb{B}\left(\rho\right)\, \right) \right],
\end{equation}
which captures the surprise that a state $\rho$ is not stationary under the process $\mathbb{B}$. The quantity $\mathbf{R}\left[\rho \,\Vert \,\mathbb{B}\, \right]$ captures the approach to quantum thermodynamic equilibrium because it never increases in time. The proof for this uses the semigroup property of map $\mathbb{B}^{(0,t)}$ and Eq.~(\ref{rentropydec}) to obtain:
\begin{align}\label{eq:relativenetropy}
\mathbf{R}\left[\,\mathbb{B}^{(0,t)}(\rho) \,\Vert \,\mathbb{B}^{(0,t)}(\mathbb{B})\, \right]&=S\left[\;\mathbb{B}^{(0,t)}(\rho)\; \Vert\; \mathbb{B}^{(0,t)}(\,\mathbb{B}(\rho)\,)\;\right]\nonumber \\ 
&\leq S\left[\,\rho\, \Vert\, \mathbb{B}(\rho)\,\right]=\mathbf{R}\left[\rho \,\Vert \,\mathbb{B}\right]. \nonumber
\end{align}
Using the property that $\mathbb{B}(\rho)$ is a stationary state of $\mathbb{B}^{(0,t)}$ such that $\mathbb{B}^{(0,t)}(\,\mathbb{B}(\rho)\,)=\mathbb{B}(\rho)$, we complete the proof:
\begin{equation}
\mathbf{R}\left[\,\mathbb{B}^{(0,t)}(\rho) \,\Vert \,\mathbb{B}\, \right]\leq\mathbf{R}\left[\rho \,\Vert \,\mathbb{B}\right].\nonumber\label{increase}
\end{equation}
This quantity captures the approach to thermodynamic equilibrium because
\begin{equation}
\lim_{t\rightarrow \infty} \mathbf{R}\left[\,\mathbb{B}^{(0,t)}(\rho) \,\Vert \,\mathbb{B}\, \right]=+0.\quad\square\nonumber
\end{equation}
When a state is in the stationary set of the dynamics, its relative entropy with respect to the bath is zero, and therefore we say it is in thermodynamic equilibrium.

\subsection{B. Entropy production rate for quantum processes}

\begin{figure}[h]\begin{center}
\includegraphics[keepaspectratio=true,width=.85\linewidth]{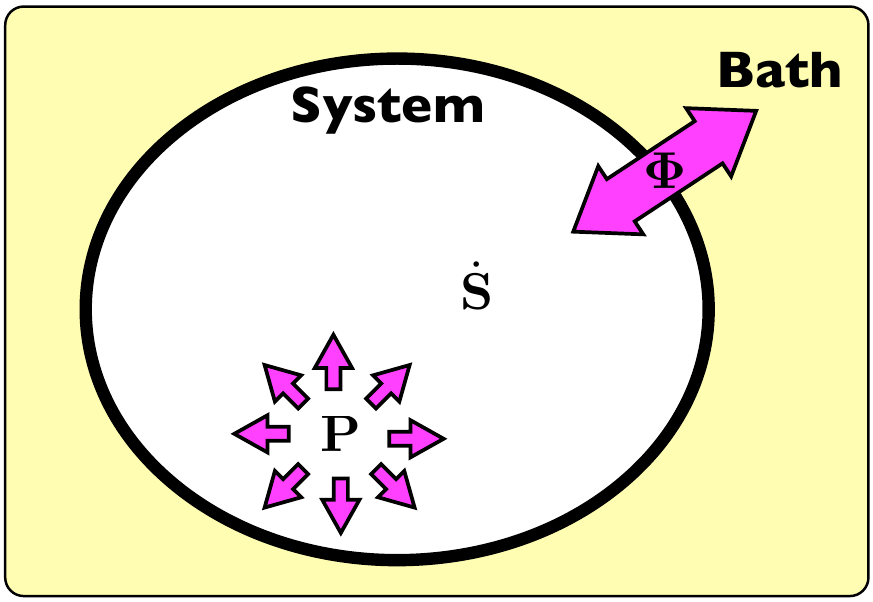}  
\caption{\emph{ Quantum entropy rate equation.} The entropy rate $\dot{\mathbf{S}}$ depends on the entropy flux to the bath $\boldsymbol{\Phi}$ and the irreversibility of the dynamics characterized by the entropy production rate $\mathbf{P}$. The second law of thermodynamics is simply $\mathbf{P}\geq0$ and is satisfied by all quantum Markovian processes, including decoherence.}
\label{fig:entropybalance}\end{center}
\end{figure}

Using the property that $\mathbb{B}(\rho)$ is stationary, we take the time derivative of Eq.~(\ref{eq:relativenetropy1}) and obtain the entropy rate equation:
\begin{equation}\label{eq:entropyrate}\begin{array}{ccc}
\underbrace{-\mbox{Tr}\left[\dot{\rho} \log \rho \right]}&=  \underbrace{\mbox{Tr}\left[\dot{\rho} \log \mathbb{B}\left(\rho\right) \right]} & \underbrace{- \frac{d}{dt}\mathbf{R}\left[ \rho\Vert\mathbb{B}\right]},\\
\dot{\mathbf{S}}=&  -\boldsymbol{\Phi} & +\mathbf{P},
\end{array} 
\end{equation}
where $\dot{\mathbf{S}}=-\mbox{Tr}\left[\dot{\rho} \log \rho \right]$ is the rate of the von Neumann entropy of the system, $\boldsymbol{\Phi}$ is the entropy flux due to the bath, and $\mathbf{P}$ is the entropy production rate due to irreversibility (see Fig.~(\ref{fig:entropybalance})). The second law of quantum thermodynamics is $\mathbf{P}=- \frac{d}{dt}\mathbf{R}\left[ \rho\Vert \mathbb{B}\right]\geq0$, which means that irreversibility cannot decrease the quantum entropy production. This has already been proven for the special case where the dynamics are relaxation $\mathbb{B}_r$ to a single Gibbs state \cite{Spohn:1978uv}.

We are interesting in showing how irreversibility due to more general quantum processes, such as decoherence, contributes to the entropy production. For this, we generalize the proof to all possible $\mathbb{B}$. We start by using the result from Eq.~(\ref{increase}) that relative entropy with respect to the process can never increase. Therefore, its time derivative,
\begin{equation}
 \frac{d}{dt}\mathbf{R}\left[ \rho\Vert\mathbb{B}\right]=\lim_{t\rightarrow+0}\frac{\mathbf{R}\left[\,\mathbb{B}^{(0,t)}(\rho) \,\Vert \,\mathbb{B}\, \right]-\mathbf{R}\left[\rho \,\Vert \,\mathbb{B}\right]}{t}\leq 0,\nonumber
\end{equation}is never positive. It follows that $\mathbf{P}\geq0$ and that the second law of thermodynamics is satisfied for any quantum process. Since this is true for any quantum process, this shows how decoherence contributes to the entropy production. Interestingly, quantum contributions to the entropy production lead to novel effects that cannot be described classically. We discuss these effects in the next sections.

 \subsection{C. Heat rate from decoherence}

The irreversible loss of quantum coherences to a decoherence bath can lead to heat rates. To show this, we now consider a master equation for decoherence dynamics of the form: 
 \begin{equation}
 \mathbb{L}_d(\rho)=\sum_j \gamma\Big(\; 2\vert j \rangle\langle j \vert\,\rho\,\vert j \rangle\langle j \vert -\vert j \rangle\langle j \vert\,\rho-\rho \,\vert j \rangle\langle j \vert\; \Big),
 \end{equation} 
which is equivalent to a continuous measurement along the basis $\{\vert j \rangle\}$.  Equilibrium in a decoherence bath can be fully characterized by this preferred basis. Such a bath has no temperature associated with it, but can create a heat rate by changing quantum coherences. For this, we will assume that, at some time, the Hamiltonian of the system is $H=\sum_k E_k\vert E_k \rangle\langle E_k \vert$, where  $\{ \vert E_k \rangle \}$ is not in the preferred basis, leading to quantum coherences of the form $\langle j \vert E_k \rangle$. From the first law of quantum thermodynamics, the heat rate due to a dephasing process is
\begin{align}
\dot{Q}_d&= \mbox{Tr}\left[\,H\;\mathbb{L}_d(\rho)\, \right],\nonumber \\
&=  \gamma\sum_k E_k \sum_j \; 2 \langle j \vert \rho \vert j \rangle \;\langle j \vert E_k \rangle\langle E_k \vert j \rangle \nonumber\\&\qquad- \langle j \vert \rho  \vert E_k \rangle\;\langle E_k \vert j \rangle - \langle E_k \vert \rho \vert j \rangle\;\langle j \vert E_k \rangle. \label{decoheat}
\end{align} 
This heat rate depends on the quantum coherence $\langle j \vert E_k \rangle$. When the coherence vanishes, so does the heat rate. For many states $\rho$ there is a heat rate, even though a decoherence bath has no unique temperature associated with it. This heat rate is an example of how quantum processes can created thermodynamic flows that depend on quantum variables.

To be more concrete, we now consider a two level system as an example, $\rho = \frac{1}{2}\left( \mathbf{I}+x\sigma_x+z\sigma_z \right)$, where $\mathbf{I}=\vert 1 \rangle\langle 1 \vert+\vert 0 \rangle\langle 0 \vert$, $\sigma_z=\vert 1 \rangle\langle 1 \vert-\vert 0 \rangle\langle 0 \vert$ and $\sigma_x=\vert 1 \rangle\langle 0 \vert-\vert 0 \rangle\langle 1 \vert$. We also assume that the preferred basis of the dephasing bath is $\{ \vert 0 \rangle,\vert 1 \rangle \}$. The system parameter $x$ is thus the amount of coherence of the system.

The Hamiltonian is at a different basis $H=E\vert E \rangle \langle E \vert$ where we chose $\vert E \rangle = \frac{1}{\sqrt{2}}\left(\,\vert 0 \rangle+\vert 1 \rangle  \,\right)$. This choice makes quantum coherences to simply be $\langle 1 \vert E \rangle=\langle 0 \vert E \rangle=\frac{1}{\sqrt{2}}$. The decoherence heat rate for this example is
\begin{widetext}
\begin{equation}\begin{array}{ccc}
\dot{Q}_d&= & \gamma E \left[ \;1- \frac{1}{\sqrt{2}}\big( \langle 1 \vert \rho \vert E \rangle+\langle 0 \vert \rho \vert E \rangle
+\langle E \vert \rho \vert 1 \rangle+\langle E \vert \rho \vert 0 \rangle
\big)\right]=\gamma E x.
\end{array} 
\end{equation} 
\end{widetext}
The quantum heat rate is proportional to amount of quantum coherence of the system $x$. This is an example of how a decoherence bath can create a heat rate that depends on quantum coherence.

Previous publications showed the dependence of a quantum Carnot engine in terms of quantum coherences \cite{Scully:2003vt,Scully:2010cn}. In those cases, the quantum coherences are subject to decoherence. We suggest the equivalent but alternative interpretation that such decoherence produces a heat rate, that has no temperature associated with it, which affects the efficiency of this engine.

Another publication has suggested how continuous quantum measurements can be used for thermodynamic control \cite{Erez:2008ep}. In their model they considered the temperature relaxation of a state while also under the influence of frequent quantum measurements. We suggest the interpretation that since such measurements can be modeled as an additional decoherence bath, the act of measuring the state introduces another source of heat. This heat rate from decoherence serves to control the temperature of the state. 

  \subsection{D. Quantum reciprocal relations}
  
  The classical Onsager reciprocal relations \cite{Onsager:1931uu,Miller:1960ci} use a linear approximation of the flows in terms of the forces to study non-equilibrium thermodynamics. They can be derived using classical stochastic processes and considering deviations from the stationary state of each bath~\cite{1976RvMP...48..571S}. This method has been very successful to understand how many classical irreversible processes affect each other~\cite{Mazur}. Important work showed that the classical Onsager relations were recovered for quantum relaxation processes \cite{Spohn:1978wt,Lendi:2003uo}. We are interested in using them to study the effects of decoherence on other thermodynamic variables. For this, we must extend these relations to more general quantum processes.
  
  When a system is coupled to many baths, the interplay between them can lead to a non-equilibrium steady state $\nu$. This state creates an entropy flux to each bath of the form $\boldsymbol{\Phi}_b=-\mbox{Tr}\left[\mathbb{L}_b(\nu) \log \mathbb{B}_b\left({\nu}\right) \right]$. The total entropy production rate is
\begin{align}\label{eq:ness1}
\mathbf{P}& =  \sum_b\mbox{Tr}\left[\;\mathbb{L}_b(\nu)\;\big(\log \mathbb{B}_b\left(\nu\right) - \log \nu\big)\;\right]\nonumber \\
& =\sum_b\mbox{Tr}\left[J_b \, X_b \right],
\end{align}
where  
\begin{equation}\label{eq:J}
J_b= \mathbb{L}_b(\nu)
\end{equation}
represents a quantum flow and 
\begin{equation}\label{eq:X}
X_b=\log \mathbb{B}_b\left(\nu\right) - \log \nu
\end{equation} 
is the corresponding quantum thermodynamic force. The rate of change represented by $J_b$ says how far away from equilibrium there is a flow to bath $b$. The force $X_b$ represents deviations of the non-equilibrium steady state from the corresponding stationary state for bath $b$.  

Quantum irreversible processes are more general than the classical irreversible processes and require density matrices to express deviations from the stationary state. This is why $J_b$ and $X_b$ are matrices. The quantum Onsager linearization has to be given by a super-operator $\mathbb{M}_{b,a}$ such $J_b\approx \sum_a \mathbb{M}_{b,a}(X_a)$. The entropy production rate in terms of quantum forces is:
\begin{equation}\label{eq:onsager}
\mathbf{P}=  \sum_{a,b}\mbox{Tr}\left[\mathbb{M}_{b,a}(X_a) \, X_b \right].
\end{equation}
We now show that the relations between these forces are reciprocal. To do this, we start with Eq.~(\ref{eq:X}), and linearly approximate $\log \mathbb{B}_a\left(\nu\right)\approx \delta \mathbb{L}_a\left(\nu\right)$ to obtain from Eq.~(\ref{eq:X}) the approximation $\delta \mathbb{L}_a\left(\nu\right)\approx \log \nu+X_a$. The Heisenberg picture representation of $\mathbb{L}_a$, $\mathbb{L}^\dagger_a$, allows us to write 
\begin{equation}\label{eq:nu}
\nu\approx \sum_a C\mathbb{L}^\dagger_a\left(\log \nu+X_a\right),
\end{equation} where C is a constant. We expand to first order $\log \nu\approx \nu-\mathbb{I}$, and Eq.~(\ref{eq:nu}) becomes $\nu\approx \sum_a C\mathbb{L}^\dagger_a\left(X_a\right)+C\mathbb{L}^\dagger_a\left(\nu\right)$,
because $\mathbb{L}^\dagger_a\left(\mathbf{I}\right)=0$. Recall that since $\nu$ is a non-equilibrium steady state, then $\sum_a\mathbb{L}_a\left(\nu \right)=\mathbb{L}\left(\nu \right)=0$. Assuming quantum detailed balance for the total dynamics $\mathbb{L}$ \cite{Gorini:1978tn,Kossakowski:1977ug,Temme:2011eo}, the Heisenberg picture operator also follows $\mathbb{L}^\dagger\left(\nu \right)=\sum_a\mathbb{L}^\dagger_a\left(\nu \right)=0$. With this, we approximate the non-equilibrium steady state in the linear regime as $\nu\approx \sum_a C\mathbb{L}^\dagger_a\left(X_a\right)$.

This equation shows how we can express the non-equilibrium steady state $\nu$ linearly in terms of the deviations from equilibrium $X_a$ from each of the baths.
From Eq.~(\ref{eq:ness1}), the flow in the linear regime is
\begin{equation}\label{flow}
J_b= \mathbb{L}_b(\nu)\approx\sum_a \mathbb{M}_{b,a}(X_a) \equiv\sum_a C\mathbb{L}_b\left(\,\mathbb{L}^\dagger_a\left(X_a\right)\,\right) ,
\end{equation}
The Heisenberg picture of the $\mathbb{M}_{b,a}$ is $\mathbb{M}^\dagger_{a,b}(\,\cdot\,)=C\mathbb{L}^\dagger_a\left(\,\mathbb{L}_b\left(\,\cdot\,\right)\,\right)$, which leads to the quantum reciprocal relations:
\begin{equation}\label{recip}
\mathbb{M}_{b,a}(\,\cdot\,)=\mathbb{M}_{a,b}^\dagger(\,\cdot\,).
\end{equation}

The quantum reciprocal relations give us a phenomenological way to understand the non-equilibrium interplay between different quantum thermodynamic baths. They are useful because they do not require knowledge of the master equation of each bath. We only need to know the deviations from equilibrium for each bath, given by the matrix Eq~(\ref{eq:X}). These can account for thermodynamic effects due to decoherence.

 \subsection{E. Reciprocal relations between decoherence and transport}
  We illustrate how the Onsager reciprocal relations can be used to understand the interplay between decoherence and transport. For this, we consider the simple device described in the main text, Fig.~(2).
  
The Hamiltonian of the system is  
\begin{equation}
H=E_L \vert L \rangle\langle L \vert + E_R \vert R \rangle\langle R \vert +\frac{V}{2}\big(\;\vert L \rangle\langle R \vert+\vert R \rangle\langle L \vert\;\big)\nonumber
\end{equation}
where, for simplicity, we set $E_L>E_R$, $E_0=0$ and assumed $V=V^*$. The system is subject to the interplay between three baths. The bath on the left is a relaxation bath, described by the master equation:

 \begin{align}
 \mathbb{L}_l(\rho)=&(1+n_L)\Big[\; 2\vert L \rangle\langle 0 \vert\,\rho\,\vert 0 \rangle\langle L \vert -\vert 0 \rangle\langle 0 \vert\,\rho-\rho \,\vert 0 \rangle\langle 0 \vert\; \Big]\nonumber \\
 &+n_L\Big[\; 2\vert 0 \rangle\langle L \vert\,\rho\,\vert L \rangle\langle 0 \vert -\vert L \rangle\langle L \vert\,\rho-\rho \,\vert L \rangle\langle L \vert\; \Big].\nonumber
 \end{align} where $n_L=1/(e^{\beta_L E_L}-1)$. The parameter $n_L$ characterizes the stationary Gibbs distribution of this bath. The bath on the right is also a relaxation bath, with master equation:
 \begin{align}\label{masterright}
 \mathbb{L}_r(\rho)=n_R\Big[\; 2\vert R \rangle\langle 0 \vert\,\rho\,\vert 0 \rangle\langle R \vert -\vert 0 \rangle\langle 0 \vert\,\rho-\rho \,\vert 0 \rangle\langle 0 \vert\; \Big]\nonumber\\+(1+n_R)\Big[ 2\vert 0 \rangle\langle R \vert\rho\vert R \rangle\langle 0 \vert -\vert R \rangle\langle R \vert\rho-\rho \vert R \rangle\langle R \vert \Big].
 \end{align} where $n_R=1/(e^{\beta_R E_R}-1)$, which characterizes the stationary distribution of the bath in the right.  This constitute a simple quantum transport device~\cite{Mohseni:2008wt,Plenio:2008ff,Rebentrost:2009tj,Solomon:2008ej,Vazquez:2012gg}. Its total evolution can be found by solving
 \begin{equation}
 \dot{\rho}=-i\left[H,\rho\right]+\mathbb{L}_l(\rho)+ \mathbb{L}_r(\rho).
 \end{equation}
At non-equilibrium steady state, these two baths create an energy flow $\dot{\mathbf{Q}}_r=-\dot{\mathbf{Q}}_l$ through the system by means of the coherence $\langle L\vert H\vert R\rangle=V$. To find this flow, we must solve the total master equation.
 
 In practice, such a device is also subject to some quantum decoherence. When decoherence is strong, the device operates in the incoherent regime, and classical transport efficiencies are recovered \cite{Manzano:2011wz}. It has been shown that in the intermediate regime between coherent and incoherent transport, the efficiency can be maximized, which is important in energy transport in photosynthesis \cite{Mohseni:2008wt,Rebentrost:2009tj}.

 To study this intermediate regime, decoherence is introduced as a third bath. Its master equation is:
 \begin{widetext}
  \begin{equation}\label{masterdeph}
 \mathbb{L}_d(\rho)=\gamma\Big(\; 2\vert L \rangle\langle L \vert\,\rho\,\vert L \rangle\langle L \vert -\vert L \rangle\langle L \vert\,\rho-\rho \,\vert L \rangle\langle L +2\vert R \rangle\langle R \vert\,\rho\,\vert R \rangle\langle R \vert -\vert R \rangle\langle R \vert\,\rho-\rho \,\vert R \rangle\langle R\vert\; \Big).
 \end{equation} 
  \end{widetext}
 The introduction of decoherence changes the non-equilibrium steady state of the device, and would require the solution of a new master equation to compute the new energy flow $\dot{\mathbf{Q}}_l=\mbox{Tr}\left[H \mathbb{L}_l(\nu)\right]$. 
 
 Instead of solving this equation, we could use the reciprocal relations to approximate this calculation in terms of deviations from equilibrium. We now use this to understand understand the relationship between decoherence and energy transport. Decoherence introduces a new heat rate into the system, as in Eq.~(\ref{decoheat}). This heat rate is given by $\dot{\mathbf{Q}}_d=\mbox{Tr}\left[H \mathbb{L}_d(\nu)\right]$. Both $\dot{\mathbf{Q}}_l$ and $\dot{\mathbf{Q}}_d$ seem to have a complicated relationship because they both depend on $\nu$, which requires in turn the calculation of the full dynamics.
 
 However, in the linear regime, using Eq.~(\ref{flow}), the  heat rates can be approximated as:
\begin{equation}
 \dot{\mathbf{Q}}_l\approx \mbox{Tr}\left[H\mathbb{M}_{l,l}(X_l)\right]+\mbox{Tr}\left[H\mathbb{M}_{l,d}(X_d)\right]+\mbox{Tr}\left[H\mathbb{M}_{l,r}(X_l)\right].\nonumber
 \end{equation}
The force $X_l$ ($X_r$) represents how for is the steady state from equilibrium with the Gibbs state that characterizes the bath $l$ ($r$). The force $X_d$ captures how far from equilibrium is the steady state $\nu$ from decoherence. It depends on the coherence of the non-equilibrium steady state as $\log \langle L\vert \nu\vert R\rangle$. This is a force due to the coherent coupling between the left and the right bath. It creates a thermodynamic flow that, by means of the matrix $\mathbb{M}^\dagger_{d,l}(H)$, affects $\dot{\mathbf{Q}}_l$. Using Eq.~(\ref{flow}) with Eq.~(\ref{masterright}) and Eq.~(\ref{masterdeph}), and by explicit calculation, we obtain that:
$\mathbb{M}^\dagger_{d,l}(H)\approx\mathbb{L}^\dagger_d\left(\,\mathbb{L}_l\left(H\right)\,\right)=\gamma n_l\left(\vert R \rangle \langle L \vert+\vert L \rangle \langle R \vert \right)$.
 Thus, the heat rate on the left $\dot{\mathbf{Q}}_l$ depends on the coherence force parameter $x_d=\log \langle R\vert \nu\vert L\rangle+\log \langle L\vert \nu\vert R\rangle$ as:
\begin{equation} \mbox{Tr}\left[\mathbb{M}^\dagger_{l,d}(H)X_d\right]\propto \left(\gamma n_l V\right) \, x_d=m_{ld} \; x_d,\nonumber
\end{equation}
This quantity depends only on the equilibrium distribution of the left bath, as given by parameter $n_l$, the coupling $V$, the decohernece rate $\gamma$ and the quantum coherent force $x_d$.

Reciprocally, we can also estimate how much the decoherence heat rate depends on the the temperature of the left bath. A similar explicit calculation tells us that the $\dot{\mathbf{Q}}_d$ depends on the force due to the left bath by:
\begin{equation} \mbox{Tr}\left[\mathbb{M}^\dagger_{l,d}(H)X_d\right]\propto \left(\gamma n_l V\right)\, x_l=m_{dl} \; x_l.\nonumber
\end{equation}
where $x_l=\langle L \vert\log\nu\vert R \rangle + \langle R \vert\log\nu\vert L \rangle$ is a force parameter due to the coherent coupling driving the non-equilibrium steady state away from equilibrium with respect to the left bath. Clearly, the flows $\dot{\mathbf{Q}}_l$ and $\dot{\mathbf{Q}}_d$ are related reciprocally by their rates because $m_{ld}=m_{dl}$. 

The power of the quantum reciprocal relations lie in their generality. Instead of a lengthy calculation depending on all the details of the master equation, to study the role of decoherence on quantum transport, we can estimate the relationship between the flow due to the bath on the left and the flow due to decoherence simply as phenomenological reciprocal relations. Even without knowledge of the master equations for each bath, or the parameters $\gamma, V, n_l$, we can still conclude that the relationship between decoherence and transport depends reciprocally between the transport force $x_l$ and the decoherence force $x_d$ by $m_{ld}=m_{dl}$. That is, decoherence affects transport exactly as much as transport phenomena affects decoherence.

\bibliography{qthermo.bib}
\small
\subsection{ Correspondence} Emails
should be addressed to C.A.R.R (cesar.rodriguez@bccms.uni-bremen.de) and A.A.-G. (aspuru@chemistry.harvard.edu).

\end{document}